\def \yskip{\penalty-50\vskip3pt plus 3pt minus 2pt}
\def \reference{\par \yskip \noindent \hangindent .4in \hangafter 1}
\def \abc#1#2#3#4 {\reference#1, {\sl#2}, {\bf#3}, #4}
\def \blank {\lower 5pt\hbox to 0.75in{\hrulefill}}

\def \cm{~\rm{cm}}
\def \s{~\rm{s}}
\def \km{~\rm{km}}

\def \g{~\rm{g}}
\def \AU{~\rm{AU}}
\def \erg{~\rm{erg}}

\def \yrs{~\rm{yrs}}
\def \yr{~\rm{yr}}

\def \lesssim{\mathrel{<\kern-1.0em\lower0.9ex\hbox{$\sim$}}}
\def \gtrsim{\mathrel{>\kern-1.0em\lower0.9ex\hbox{$\sim$}}}

\documentstyle[11pt,aasms,tighten,flushrt]{article}

\begin{document}
\small

\setcounter{page}{1}

\begin{center}
\bf
THE DEPARTURE OF $\eta$ CARINAE FROM AXISYMMETRY \\
AND THE BINARY HYPOTHESIS
\end{center}

\begin{center}
Noam Soker\\
Department of Physics, University of Haifa at Oranim\\
Oranim, Tivon 36006, ISRAEL \\
soker@physics.technion.ac.il 
\end{center}


\begin{center}
\bf ABSTRACT
\end{center}

 I argue that the large scale departure from axisymmetry of the $\eta$
Carinae nebula can be explained by the binary stars model
of $\eta$ Carinae.
 The companion diverts the wind blown by the primary star,
by accreting from the wind and possibly by blowing its own
collimated fast wind (CFW).
 The effect of these processes depends on the orbital separation,
hence on the orbital phase of the eccentric orbit.
  The variation of the mass outflow from the binary system
with the orbital phase leads to a large-scale departure from 
axisymmetry along the equatorial plane, as is observed in $\eta$ Car.
 I further speculated that such a companion may have accreted a large
fraction of the mass that was expelled in the Great Eruption of
1850 and the Lesser Eruption of 1890.
 The accretion process was likely to form an accretion disk, with the
formation of a CFW, or jets, on the two sides of the accretion disk.
 The CFW may have played a crucial role in the formation of the two
lobes. 


\clearpage

\section{INTRODUCTION}

 One of the open questions regarding the massive star $\eta$ Carinae
and its nebulosity is whether the nucleus is a single or a binary
stars system. 
 The main argument used by the binary model supporters is the
$P=2020 \pm 5~$days$~ =5.5 \yr$ periodicity of the
spectroscopic event$-$fading of high excitation lines
(Damineli 1996; Damineli {\it et al.} 2000, and references therein). 
 There are other, though weaker, supporting observations in favor
of a massive companion.
 Ishibashi {\it et al.} (1999) find that the X-ray emission, which
follows the $5.5 \yr$ periodic variation, can in principle be
explained by colliding winds, where the two winds are blown
by the two components of a binary system.
 However, not everyone supports the binary model.
 Smith {\it et al.} (2000) attribute the $5.5 \yr$ periodic variation
to variation in the ionizing hard-UV flux reaching the equatorial
gas, and argue that presently there is no advantage to a binary
model over a stellar inherent instability model
(e.g., Stothers 2000) to account for the periodicity.
 In particular they argue against the suggestion raised by Damineli
{\it et al.} (2000) that the ionizing flux is emitted by the companion,
and that the periodic variation is caused by absorption of the
companion's radiation by the primary stellar wind
during periastron passages.
 Davidson {\it et al.} (2000) claim that it is not clear yet whether
$\eta$ Car is a binary system, but if it is the parameters
of the companion and orbit are not those found by Damineli (2000),
and presently remain unknown
{{{ (some new parameters are suggested by Corcoran {\it et al.} 2001). }}}

 In the present paper I argue that the departure of the nebula
around $\eta$ Car from axisymmetry can be explained by
the central close binary model.
 By departure from axisymmetry I refer to a large-scale departure,
and do not consider small blobs, filaments, and other small-scale
features.
 In that respect a binary companion in an eccentric orbit
offers an answer to question 14 raised by
Davidson \& Humphreys (1997; hereafter DH97):
 ``Why was the eruption azimuthally asymmetric; . . . ''
(they refer also to the small blobs, but here I refer only to the
large-scale asymmetry).
 In $\S 2$ I describe the departure from axisymmetry of
the nebula around $\eta$ Car.
 I then demonstrate that the parameters suggested for the binary
system can in principle lead to the formation of a nebula possessing
a large-scale departure from axisymmetry.
 In $\S 3$ I discuss some implications of the binary model for
the formation of the bipolar structure of the Homunculus and the
dense equatorial flow.
 I summarize in $\S 4$.

\section{DEPARTURE FROM AXISYMMETRY}

 Although the Homunculus seems quite axisymmetrical,
it is not perfectly so. 
 Morse {\it et al.} (1998) argue that ``the lobes exhibit
a slight banana-shaped symmetry . . ''.
The ``banana-shaped'' structure is more prominent in the velocity maps
presented by Allen \& Hillier (1993), which show the structure in 
planes parallel to the line of sight.
 These maps show that the departure of the Homunculus from axisymmetry
is mainly along the line of sight, hence hard to detect in
simple imaging of $\eta$ Car. 
 Many other structural features show a highly prominent departure
from axisymmetry in the plane of the sky.
 Figure 3 of Morse {\it et al.} (1998, see their Erratum for high
quality images), shows a clear departure from axisymmetry 
on the outskirts, $\sim 10^{\prime \prime}$ from the nucleus.
On the south-west side of the equatorial plane there is a dense
arc of gas, termed S-Ridge, while on the north-east side there is
no such an arc, but rather the ``jet'' and the ``NN bow''.
 It is clear that this departure from axisymmetry along
the equatorial plane has a large-scale structure, and can't be
attributed to instabilities in the flow or in the mass loss process. 
{{{ A same sense of asymmetry is seen in the radial velocities
map presented by Weis, Duschl \& Bomans (2001).
 On the south-west side the measured radial velocities are higher
than those on the north-east side. 
A large scale departure from axisymmetry is clearly seen also in recent
X-ray images (Weis {\it et al.} 2001; Seward {\it et al.} 2001). }}}
 Other departures from axisymmetry along the equatorial plane
are seen much closer to the nucleus.
 The $10 \mu$m image of Morris {\it et al.} (1999) shows that the peak
on the north-east side of the equatorial plane is much stronger
than the emission on the south-west side.
 The same sense of departure from axisymmetry is seen at shorter IR
bands (e.g., fig. 5 of Smith, Gehrz \& Krautter 1998 and
fig. 1 of Smith \& Gehrz 2000).
 There are indications for displacement from axisymmetry in the
equatorial plane along the south-east to north-west direction
as well, e.g., figure 3 of Smith {\it et al.} (1998) shows the
two sides of the equatorial plane to be bent toward the south-east.

 Soker, Rappaport, \& Harpaz (1998; hereafter SRH) demonstrate via
analytical calculations that when a companion is close enough to
influence the mass-loss process from an evolved star and/or from
the binary system as a whole, and the eccentricity is substantial,
the nebula around the mass-losing star will acquire a
large-scale departure from pure axisymmetry.
 An essential ingredient is that the mass-loss rate and/or geometry
varies systematically with orbital phase, due to the periodic change
in the orbital separation.
 SRH examine the displacement of the central
star from the center of the nebula, e.g., as in the planetary
nebula Hu 2-9 (Miranda {\it et al.} 2000), but the departure
from axisymmetry can manifest itself in other ways, e.g.,
one side will contain a denser section
(Soker \& Rappaport 2001).
 SRH consider two effects of the companion on the mass-loss process:
a tidal enhancement of the stellar wind near periastron, and a
cessation of the stellar wind when the Roche lobe of a mass-losing
asymptotic giant branch (AGB) star encroaches on its extended
atmosphere near periastron passage.
{{{  With regard to the first mechanism, in a recent paper Corcoran
{\it et al.} (2001) argue, based on the X-ray light curve,
that the mass loss rate from $\eta$ Car increases by a factor of
20 following periastron passage. }}}
 Soker \& Rappaport (2001) consider other processes by which a
companion can influence the mass-loss process from the system; 
the direct gravitational influence on the wind
(Mastrodemos \& Morris 1999), and the formation of a collimated
fast wind (CFW) by the companion
(Morris 1987; Soker \& Rappaport 2000; hereafter SR00).
 In the later process the companion is assumed to accrete from the
mass-losing star, to form an accretion disk, and to blow a CFW.
 The interaction between the CFW, if strong enough, and the AGB
wind will form a bipolar planetary nebula (Morris 1987; SR00).
 Another process relevant to the binary system proposed for 
$\eta$ Car is the interaction of the stellar wind blown by
the companion with the wind blown by the primary. 

 I now show that a companion star to $\eta$ Car can strongly
modulate the mass-loss process from the binary system along its
orbital motion, naturally leading to the formation of a nebula
which possesses a large-scale departure from axisymmetry. 
 I scale the binary parameters by values which were quoted in
recent years (e.g., Ishibashi {\it et al.} 1999; Damineli {\it et al.}
2000; {{{ Corcoran {\it et al.} 2001): }}}
for the mass of the primary (mass-losing) star and eccentricity I take
$M_1=80 M_\odot$ and $e=0.8$, respectively.
 For the present primary's wind I take a mass-loss rate of
 $\dot M_1=3 \times 10^{-4} M_\odot \yr ^{-1}$ and a velocity of
$v_1=500 \km \s^{-1}$.
 For the companion mass I take $M_2=30 M_\odot$
and for the companion's wind I take a
mass-loss rate of $\dot M_2=3 \times 10^{-6} M_\odot \yr^{-1}$, and a
velocity of $v_2=2000 \km \s^{-1}$.
 The semimajor axis is $a=15 \AU$.
 The proposed mechanism for the departure from axisymmetry,
including the calculations below, is applicable to a more massive model
of $\eta$ Car, as suggested by DH97, who argue that the initial and
present masses of $\eta$ Car are $160 M_\odot$ and $120 M_\odot$,
respectively.
 The accretion radius of the companion, for accretion from
the primary's wind, is
\begin{equation}
R_a \simeq {{2 GM_2}\over {v_1^2}} =
0.2 
\left({M_2} \over {30 M_\odot} \right)
\left({{v_1} \over {500 \km \s^{-1}}} \right)^{-2}  \AU.
\end{equation}
 The distance $D_2$ of the stagnation point of the colliding
winds from the companion along the line between the stars is
given by equating the ram pressures of the two winds $\rho v^2$.
 For spherically symmetric winds 
\begin{equation}
D_2= r \beta (1+\beta)^{-1} \simeq r \beta =
a \left( {{1-e^2} \over {1+e \cos \theta}} \right)
\left( {\dot M_2 v_2} \over {\dot M_1 v_1} \right)^{1/2}, 
\end{equation}
where $r$ is the orbital separation, $\theta$ is the angular
distance along the orbit ($\theta=0$ at periastron), and 
$\beta \equiv [(\dot M_2 v_2)/(\dot M_1 v_1)]^{1/2}$. 
 In the second equality I assumed $\beta \ll 1$.
 The stagnation point should be compared with the accretion radius.
 For the present winds' parameters used above $\beta \simeq 0.2$
and at periastron $r=a (1-e) = 3 \AU$, so that $D_2= 0.5 \AU$.
 The accretion radius is smaller than $D_2$, hence no accretion
will take place.
 Also, the accretion radius is quite small compared with the
orbital separation even at periastron, hence the companion
will not influence the mass-loss process much. 

 If, however, during a mass loss episode the primary's wind
velocity decreases to $v_{1} \lesssim 300 \km \s^{-1}$,
as may be suggested by some condensations along the orbital
plane of $\eta$ Car (Davidson {\it et al.} 1997;
Smith \& Gehrz 1998), the accretion radius
will be $R_a \gtrsim 0.6 \AU$, and the companion will
deflect a substantial portion of the primary's wind at periastron,
since $R_a \gtrsim 0.2 r$.
 For the parameter chosen above and $v_1=300 \km \s^{-1}$, I find
that at periastron $D_2 = 0.65 \AU \sim R_a$, hence some
accretion may occur, but only near periastron.
 At other orbital phases $D_2>R_a$, and no accretion to the companion
will take place.
  If in addition to the slower equatorial flow the mass-loss rate
is much higher,  $\dot M_1 = 0.1 M_\odot \yr ^{-1}$ as suggested for
the eruption of 1850 (the Great Eruption) that formed the lobes (DH97),
then $\beta \simeq 0.01$ and even at apastron
$D_2 \simeq r \beta = 0.4 \AU < R_a$,
hence significant accretion will occur during the entire orbital motion.
 For a mass loss rate of $\dot M_1=0.1 M_\odot \yr^{-1}$
and wind velocity of $v_1= 500 \km \s^{-1}$ the accretion 
rate by the companion at periastron, without wind disruption, is 
$\dot M_{\rm acc} \simeq \dot M_1 R_a^2/4 r^2 \simeq
10^{-4} M_\odot \yr^{-1}$.
At other orbital phases the accretion rate is lower, and may cease
near apastron.
 This may be enough for the companion to blow a CFW
(SR00), with a varying strength along its orbit.
 However, it seems (see next section) that during the
Great Eruption of 1850 and the Lesser Eruption of 1890, the equatorial
mass flux was higher than the average, and the velocity much lower,
making the accretion rate by the companion much higher,
and the proposed mechanism for causing departure from axisymmetry
much more efficient. 

 For an illustration, I assume that near periastron ($\theta=0$,
$r_p= 3 \AU$) the accretion rate from the equatorial flow
is very high, preventing any mass loss from the system when the
orbital separation is $r<4 \AU$.
 This is case 2 of mass-loss process considered by SRH.
 For $a=15 \AU$ and $e=0.8$, this corresponds to no wind being
 blown from the binary system during orbital phases of
$\vert \theta \vert \lesssim 65^\circ$.
 From the left panel of figure 2 of SRH I find that for these parameters
$<\dot y > / \Omega_K a =0.1$,
where $<\dot y>$ is the average speed of the outflowing matter
in the equatorial plane (SRH eq. 5), and $\Omega_K$ is the Kepler
frequency.
 Here $\Omega_K = 2 \pi/5.5 \yr^{-1}$ and $\Omega_K a = 81 \km \s^{-1}$,
from which I find $<\dot y > =8 \km \s^{-1}$.
The offset of the nucleus from the center of the equatorial flow
is (SRH eq. 7) $\delta= <\dot y > /v_1$, which
for a slow equatorial flow of $v_1 = 50 \km \s^{-1}$
(Davidson {\it et al.} 1997) gives $\delta \simeq 0.15$.
 I argue that this explains the departure from axisymmetry
of the equatorial ejecta near the nucleus of $\eta$ Car.
 It should be noted that for a slower flow near the binary system,
the accretion rate will be higher near {\it apastron} rather
than near periastron passages (see next section).
 As can be noticed from the right panel of figure 2 of SRH, this will
lead to a much larger departure from axisymmetry.
 A CFW blown by the accreting companion may also increase
the departure from axisymmetry.

 Finally, we note the following mechanism to cause departure from
axisymmetry, which can operate even for a circular orbit.
 If there is an eruption which lasts for a time much shorter than
the orbital period, the wind will be blown while the mass-losing
star is moving in a specific direction along its orbital motion.
 This will cause the center of the structure formed by this
impulsive mass loss to be displaced from the central binary system
by  $d=(v_o/v_1)R_n$, where $v_o$ is the orbital velocity of the
mass-losing star around the center of mass at the moment of
mass loss, $v_1$ is the expansion velocity of the mass being blown,
and $R_n$ is the distance of the ejecta from the binary system
(increasing with time).
 For the binary system consider here, the primary orbital velocity
changes from $27 \km \s^{-1}$ at apastron
to $242 \km \s^{-1}$ at periastron.
 For $v_1=500 \km \s^{-1}$ we find for this pure impulsive
mass-loss episode that $d/R_n$ can be in the range of $0.05-0.5$. 
 Of course, we do not expect such a mass-loss event, though
it is possible that the mass-loss rate has increased substantially
during a time shorter than the orbital period, say 2 years.
It will then collide with previously ejected mass and mass blown
later to form a more complicated structure.
 The overall departure will be less than for a pure impulsive
mass-loss episode $d/R_n < v_o/v_1$, but still may be noticeable if
it has occurred not too close to apastron passage, if it occurs for a
short time, and if the increase in the mass-loss rate during
the impulsive mass-loss episode is significant.

 The conclusion from this section is that for typical parameters used
by the binary model proponents, the secondary can have an 
influence on the mass-loss process which varies with orbital phase,
in particular if the primary's wind velocity is
$v_1 \lesssim 300 \km \s^{-1}$.
 This may naturally explain the large-scale departure from
axisymmetry observed in some structural features of $\eta$ Car.
 Although the arguments presented here suggest that a binary companion
can explain in principle the departure from axisymmetry,  I can't predict 
the exact shape and degree of departure from axisymmetry. 
 For this 3D gasdynamical numerical simulations are required. 

\section{IMPLICATIONS OF THE BINARY MODEL}

 In the previous section I demonstrated that a large accretion rate 
was likely to have occurred during the eruption of 1850, and possibly 
during that of 1890. 
 What is the effect of such an event?
 The answer depends strongly on the accretion rate, which itself is
very sensitive to the relative velocity between the accreting body
and the wind.
The wind could be moving very slowly at a distance of several AU
in an ``extended envelope'', which was formed during the eruption,
with the ``photosphere'' almost as big as the orbit of Saturn (DH97).
  I now speculate that the accretion rate by the companion was high,
and the companion blew a collimated fast wind (CFW) which led to
the formation of the bipolar shape.
 Bipolar symbiotic nebulae, similar in many properties to the
bipolar shape of $\eta$ Car, are known to result from binary
interaction (e.g., Corradi {\it et al.} 2000), and so is the common
view regarding the formation of bipolar planetary nebulae (SR00).

 First let me point to a difficulty  with the energy budget in models
which assume a spherical mass ejection in the 1850 eruption
(e.g., Frank, Balick, \& Davidson 1995).
 For a total ejected mass of $2.5 M_\odot$ and with an
initial velocity of $\sim 650 \km \s^{-1}$, which is the current
expansion velocity of the lobes (Smith \& Gehrz 1998),
the total kinetic energy of the ejected gas is
$E_{ks} \simeq 10^{49} \erg$.
 This is $\sim 1/3$ of the total energy radiated during the Great
Eruption of $\eta$ Car, $E_r=3 \times 10^{49} \erg$ (DH97).
{{{ Such a high efficiency of radiation to kinetic energy conversion
can occur in an explosion.
However, the duration of the Great Eruption was much longer than the
dynamical time scale (see below), hence it wasn't a regular explosion.
 Shaviv (2000) proposes a model to explain the supper-Eddington luminosity
during the Great Eruption, where some of the radiation escape while
exerting a smaller average force on matter.
This seems to reduce the efficiency of energy transfer. }}}
 The sum of the absolute values of the momentum along all directions
is $p_{s}= 3 \times 10^{41} \g \cm \s^{-1}$, whereas the
total momentum that can be supplied by the radiation during the eruption is
$p_r = \zeta E_r/c = 10^{39} \zeta \g \cm \s^{-1}$, where $\zeta$
is the average number of times a photon is scattered by the ejected
gas.
 To account for the wind's momentum, on average each photon must
be scattered $\sim 300$ times, which again means a very efficient
acceleration mechanism.
 This can be compared to another intensive mass-loss process,
but in AGB stars.
 Progenitors of most planetary nebulae terminate the AGB phase with an
intensive mass-loss phase, the ``superwind'', which lasts
$\sim 1-2\times 10^3 \yrs$.
 This time is several hundred times the Keplerian orbital time along
the AGB stellar equator, e.g., for $M_\ast= 0.6 M_\odot$ and
$R_\ast= 2 \AU$ the Keplerian orbital time is 3.7 years.
The same ratio holds for the Great Eruption of $\eta$ Car,
which lasted 20 years, $\sim 500$ times the Keplerian orbital time
on the surface of a $80 M_\odot$ star with a radius of
$R_\ast=0.5 \AU$. 
 From observations it is found (e.g., Knapp 1986) that in most cases
the momentum flux in the superwind is $\lesssim 3$ times the
momentum flux in the stellar radiation, i.e., $\zeta \lesssim 3$.
 In a minority of the cases with a higher momentum flux, dynamical
effects due to a binary companion probably play some role.
 But in all cases the total kinetic energy in the superwind is
much smaller, by a factor of $>100$, than the total radiated
energy in the same period of time.
 The present momentum flux in the wind of $\eta$ Car is
only $20 \%$ of the radiation momentum flux (White {\it et al.} 1994),
and therefore the wind can be explained by radiation pressure.
{{{ From this discussion it seems that it is possible to
explain the kinetic energy of the Grat Eruption with a single
star model (e.g., Shaviv 2000), but a very efficient acceleration
mechanism is required.
As I suggest below, an accreting binary companion can supply some
of the kinetic energy.}}}
 
 The present kinetic energy of the lobes is much lower than that
required in a spherical eruption.
 Assume for simplicity spherical lobes, i.e., a shape of
$r=2r_0 \sin \phi$, where $\phi$ is the angle from the
equatorial plane, $r$ is the distance from the nucleus,
and $2r_0 \simeq 3 \times 10^{17} \cm$ is the
diameter of each lobe (DH97).
 I assume for simplicity that most of the mass is on the outer
boundary of the lobes, and
that each mass element is expanding at a constant velocity
since eruption, so that the velocity as a function of angle from the
equatorial plane is
\begin{equation}
v_w(\phi)= 650 \sin \phi \km \s^{-1}.
\end{equation}
 For the distribution of mass with the angle $\phi$ I take a simple
form for the mass density per unit solid angle, defined from the center
of $\eta$ Car (not from the centers of each of the spherical lobes), 
\begin{equation}
 m = m_0 (1-K \sin^\gamma \phi),
\end{equation}
where $m_0$, $K$ and $\gamma$ are constants. 
 For a constant mass per unit solid angle $K=0$, whereas for a
 concentration of mass toward the equatorial plane $0< K \leq 1$.
 We can integrate for the kinetic energy
$E_{kns}=\int (m v_w^2/2) 2 \pi \cos \phi d \phi$
and for the total mass in the lobes
$M_{ns}=\int 2 \pi m \cos \phi d \phi$.
 Evaluating the integrals gives the total kinetic energy in the lobes
under these assumption as
\begin{equation}
E_{kns} =
10^{49}
\left(1-{{3K}\over{3+\gamma}} \right)
\left(3-{{3K}\over{1+\gamma}} \right)^{-1}
\left( {{M_{ns}} \over {2.5 M_\odot}} \right)
\left( {{v_{po}} \over {650 \km \s^{-1}}} \right)^2 \erg,
\end{equation}
where $v_{po}$ is the wind velocity along the polar directions.
 For a constant mass per unit solid angle $K=0$, and the kinetic energy
of the ejected mass is a third of that in a spherical ejection
$E_{kns}=(1/3)E_{ks}$.
 The case where $2/3$, instead of $1/2$, of the total mass is within
$\vert \phi \vert < 30^\circ$ and $\gamma=1$, has $K=0.8$.
In that case the kinetic energy of the ejected mass is
$E_{kns}=(0.22) E_{ks}$.
 Since more mass is actually concentrated toward the equator (DH97),
the kinetic energy is even lower, and we can safely take here
$E_{kns} \simeq  0.1-0.2 E_{ks} \simeq 1-2\times 10^{48} \erg$.
 This means that models in which the fast ejecta is mainly
along the polar directions (e.g., Frank, Ryu \& Davidson 1998), require
an order of magnitude less energy than models with spherical ejection.
 In the binary model the CFW (or jets) along the polar directions is
(are) blown by the companion (SR00).
 The CFW can form a hot bubble inside each lobe, and efficiently
accelerate the slowly moving gas, ejected by the mass-losing star,
to higher velocities (SR00).

 I now examine the feasibility of such a scenario.
  A rotating star close to the Eddington luminosity limit will
form a slow equatorial flow (Maeder \& Meynet 2000), having
an expansion velocity of the order of the rotation velocity, which
will be slower than the orbital velocity along most of the orbit
of the companion.
 Zethson {\it et al.} (1999) have detected slowly expanding equatorial
gas, which they claim originated hundreds of years before the Great
Eruption of 1850,  {{{ although fast equatorial ejecta
exit as well (Morse {\it et al.} 1998). }}}
 Substituting the Keplerian velocity in the expression for the accretion
radius of the companion (eq. 1) gives, for the ratio of the accretion
radius to the orbital separation $r=a(1-e^2)/(1+e \cos \theta)$,
\begin{equation}
{{R_a}\over{r}} =
\left( {{ 2 M_2} \over {M_1+M_2}} \right)
\left( {{1+e \cos \theta} \over {1+2e \cos \theta + e^2}} \right).
\end{equation}
 For the parameters used in the previous section,
 $M_1=80 M_\odot$, $M_2=30 M_\odot$, and $e=0.8$, I find 
$R_a/r=0.3$ at periastron ($\cos \theta =1$),
and $R_a>r$ at all phases where $\cos \theta < - 0.9$.
 This large $R_a/r$ ratio means that the companion in such a system
accretes a large fraction of the slowly expanding equatorial flow.
 The density at the location of the companion can be much higher than
that expected for a pure wind, especially near periastron,
since some of the material in the extended envelope may fall back on
the primary star.
 This means that the total accreted mass maybe much larger than
that expelled during the eruption.
 I scale the mass blown in the CFW with $M_c = 0.25 M_\odot$,
which is the case if the accreted mass is equal to the ejected mass of
$2.5 M_\odot$, and a fraction $0.1$ of it is blown in the CFW
(or jets), and the CFW speed is scaled by the escape velocity
$v_c \simeq 2000 \km \s^{-1}$ from the companion.
 The total kinetic energy of the CFW is
$E_c=10^{49} (M_c/0.25 M_\odot)
(v_c/2000 \km \s^{-1})^2 \erg$.
 This is more than the required energy in the non-spherical mass
ejection to form the lobes, as mentioned above.

 That a fast mass loss along the polar directions can form the
desired morphology was demonstrated by numerical simulations
performed by Frank {\it et al.} (1998), although their idea
was of a single star, whereas in the present scenario the companion
blows the CFW simultaneously with the eruption of the primary star
(SR00).
 This scenario, like that of Frank {\it et al.} (1998), avoids the
problems of the interacting winds model, some of which are
summarized by Dwarkadas \& Balick (1998).
  One of the problems mentioned by Dwarkadas \& Balick
is that the massive disk required in the interacting wind model
to confine the spherical ejection is not found.
 The claim by Morris {\it et al.} (1999) for the presence of a
massive,  $\sim 15 M_\odot$, torus of cold gas around the nucleus
of $\eta$ Car is disputed by Davidson \& Smith (2000), who claim that
a correct model gives an equatorial gas distribution which is
``incapable of generating the pinched waist''.
 Dwarkadas \& Balick (1999) proposed instead
that the spherically ejected mass (during the eruption)
interacts with a very dense torus at several AU around the nucleus.
  The main problem in this scenario, in addition to the energy and
 momentum budget problem mentioned above, is the formation of a dense
torus around and close to the nucleus.
  A massive companion can play a significant role here
(Mastrodemos \& Morris 1999).

 Finally we note the high velocity gas ejected in the equatorial plane.
 The flow in the equatorial plane is very complicated, with both slowly
expanding, $v \sim 50 \km \s^{-1}$ gas
(e.g., Zethson {\it et al.} 1999); and fast moving,
velocity of $\sim 300 \km \s^{-1}$, features (e.g., Smith \& Gehrz 1998).
 A large fraction of the equatorial gas was ejected in the
Lesser Eruption of 1890 (Davidson {\it et al.} 1997;
Smith \& Gehrz 1998), rather than during the Great Eruption of 1850.
 The equatorial ejecta possesses departure from axisymmetry,
as noted in the previous section.
 As noted by SR00, when the momentum flux of the CFW (or jets) blown by
the companion is much smaller than the momentum flux of the primary's
wind, the CFW will be strongly bent so that it will flow close to the
equatorial plane.
 Based on that, I suggest that some of the fast moving gas in the
equatorial plane was ejected by the companion at high speed,
but because of the relative (to the primary's wind) low momentum
flux of the CFW it was bent toward the equatorial plane.
 The ratio of the momentum fluxes of the primary's wind and CFW
depends mainly on the accretion rate and primary's wind concentration
toward the equator.
 It is possible that during the Great Eruption of 1850 the conditions
were favorable for the formation of a very strong CFW (e.g., slowly
expanding wind concentrated toward the equatorial plane), which
forms the two lobes, whereas during the Lesser Eruption of 1890, only
a weak CFW was formed, but still strong enough to form fast moving
gas in the equatorial plane.

\section{SUMMARY}

In the present paper I argue that the large-scale departure from
axisymmetry of the $\eta$ Carinae nebula can be explained by
the binary nucleus model.
 Using binary parameters as quoted by the binary supporters, 
 I found that the companion was likely to substantially influence the
mass-loss process from the binary system.
 The degree by which such a companion diverts the outflow depends on
the orbital separation, hence on the orbital phase in the 
eccentric orbit.
  The modulation of the mass loss process with the orbital phase
may lead to a detectable departure from axisymmetry (SRH),
as is observed in $\eta$ Car.

I speculated that if such a companion exists, it may have
accreted a large fraction of the mass that was expelled in the
Great Eruption of 1850 and the Lesser Eruption of 1890.
 This requires that the matter in the equatorial plane was moving very
slowly, at $\sim 50 \km \s^{-1}$, during these eruptions.
 The accretion process was likely to form an accretion disk, with the
formation of a collimated fast wind (CFW), or jets, on the
two sides of the accretion disk.
  I showed that a CFW of $\sim 0.25 M_\odot$, which could be formed
if the accreted mass was equal to the mass that was blown
into the lobes in the Great Eruption, $2.5 M_\odot$, and $\sim 10 \%$
of it was blown into the CFW, which was blown at $2,000 \km \s^{-1}$,
can account for the total kinetic energy of lobes of $\eta$ Car.
 The CFW, therefore, was likely to be a significant factor in
shaping the lobes of $\eta$ Car.

  If the CFW blown by the companion is weak, i.e., its momentum flux
is small, it will be sharply bent by the slow wind blown by the primary
star.
 The CFW will flow parallel to the equatorial plane, leading to fast
outflowing material near the equatorial plane.
 I therefore speculated that during the Lesser Eruption of 1890 the CFW
was indeed weak, leading to the formation of the fast
equatorial outflow which was expelled then.

\bigskip

{\bf ACKNOWLEDGMENTS:}
{{{  I thank an anonymous referee for helpful comments. }}}
 This research was supported in part by grants from the
US-Israel Binational Science Foundation.


\end{document}